\documentclass[aps,pra,preprint,superscriptaddress,showpacs]{revtex4-1}
\usepackage[utf8]{inputenc}
\usepackage{graphicx}
\usepackage{amsmath}
\usepackage{bm}
\usepackage{color}

\usepackage[normalem]{ulem}

\newcommand{\bn}[1]{\mbox{\boldmath $#1$}}

\begin{document}

%\preprint{}

\title{Valence-band effective-potential evolution for coupled holes}

\author{J.~J.~Flores-Godoy}
\email{job.flores@uia.mx}
%\homepage[]{Your web page}
%\thanks{}
%\altaffiliation{}
\affiliation{Dep. de F\'{\i}sica y Matem\'{a}ticas, Universidad Iberoamericana, C.P. 01219, M\'exico D. F.}

\author{A.~Mendoza-\'{A}lvarez}
\email{alejandro.mendoza@uia.mx}
%\homepage[]{Your web page}
%\thanks{}
%\altaffiliation{}
\affiliation{Dep. de F\'{\i}sica y Matem\'{a}ticas, Universidad Iberoamericana, C.P. 01219, M\'exico D. F.}

\author{L.~Diago-Cisneros}
\email{ldiago@fisica.uh.cu}
%\homepage[]{Your web page}
%\thanks{}
%\altaffiliation{}
\affiliation{Facultad de F\'{\i}sica. Universidad de La Habana, C.P.10400, La Habana, Cuba.}
\affiliation{Dep. de F\'{\i}sica y Matem\'{a}ticas, Universidad Iberoamericana, C.P. 01219, M\'exico D. F.}

\author{G.~Fern\'{a}ndez-Anaya}
\email{guillermo.fernandez@uia.mx}
%\homepage[]{Your web page}
%\thanks{}
%\altaffiliation{}
\affiliation{Dep. de F\'{\i}sica y Matem\'{a}ticas, Universidad Iberoamericana, C.P. 01219, M\'exico D. F.}

%\collaboration{}
%\noaffiliation

%\date{}

\begin{abstract}
   We present the metamorphosis in the effective-potential profile of layered heterostructures, for several III-V semiconductor binary compounds, when the band mixing of light and heavy holes increases. A root-locus-like procedure, is directly applied to an eigenvalue quadratic problem obtained from a multichannel system of coupled modes, in the context of multiband effective mass approximation. By letting grow valence-band mixing, it is shown the standard fixed-height rectangular potential-energy for the scatterer distribution, to be a reliable test-run input for heavy holes. On the contrary, this scheme is no longer valid for light holes and a mutable effective \emph{band offset} profile has to be considered instead, whenever the in-plane kinetic energy changes.
\end{abstract}

%\keywords{}

\maketitle

\section{Introductory Outlines}
\label{sec:Intro}
 {For holes, the single-band effective mass approximation becomes inappropriate for describing the quantum properties of band-mixed states.} These quasi-particles are essentially mixed even far away from the scattering potential we are interested in. Many theoretical models have been developed to face this problem in multi-layered systems. In this work, we deal with the $\bn{k\cdot p}$ two-band effective-mass Kohn-L\"uttinger  {(KL)} model \cite{KL55}, due to its widely accepted accuracy for describing dynamics of elementary excitations and electronic properties in the valence band \cite{Ikonic92a,Ikonic92b}.

We  {have} focused the multiband-hole band mixing-phenomenon itself due to the strong dependency of hole quantum transport physics upon the wave vector transversal to the main direction of transmission ($\kappa_{\textsc t}$). This phenomenology, early quoted by Wessel and Altarelli in resonant tunneling \cite{Wessel89}, has been lately stressed for real-life technological devices \cite{Klicmeck01}. Fundamental condensed-matter studies \cite{Milanovic88,RF04}, had propelled us into the present modelling, since they have predicted the modification of the effective potential in the electronic case. Here the key lies in the fact, that the potential-energy profile distribution in either of the binary-alloy slab might evolve, depending on the value of the transversal component of the wave vector \cite{Milanovic88,RF04}. With respect to the electrons quantum transport through layered heterostructures, the situation become more cumbersome, whenever the holes are involved. These last charge carriers  {---}been  {t}he heaviest ones {---}, notably dictate the threshold response of technological devices.

 {We recall earlier contributions to the hole subband structure in quantum wells, that has been previously reported within the envelope-function approximation for KL Hamiltonians} \cite{Ikonic92a,Ikonic92b}.  {The hole-state quantization calculated in truncated parabolic confining potential} \cite{Ikonic92a},  {as well as for different grown directions} \cite{Ikonic92b},  {were undoubtedly striking results, whose validity was amply discussed in the literature. Having focused mainly the hole spectrum in quantum systems, they do not address, for example, the problem of the effective potential metamorphosis that undergo the mixing of holes.} {While these studies} \cite{Ikonic92a,Ikonic92b}  {have added substantial contributions to the elucidation of the valence-subband structure in quantum systems, there remain some aspects which do not appear to have received yet sufficient attention and/or because of their interest deserve further clarification.}  {This is essentially the case of the carriers' transverse motion influence on the effective scattering potential they interact with.}  {This is crucial for quantum transport calculation, a question soon to be considered partially in this paper.}

On general grounds, for $\kappa_{\textsc t} \approx 0$ the  \emph{band offset} of the effective potential $V_{\mathrm{eff}}$ is given by the difference for $3D$ band edge levels \cite{LCRP06,RF04}. {By letting $\kappa_{\textsc t}$ grow}, the band mixing effects arise and the effective \emph{band offset} changes. A comprehensive analysis of this subject, describes largest reduction for the piecewise constant $V_{\mathrm{eff}}$ height as a function of $\kappa_{\textsc t}$, for light holes (\emph{lh}) respect to that for heavy holes (\emph{hh}) \cite{LCRP06}. Several new and {peculiar} features for each hole flavor  {have} been found, and they enhance the novelty of the present study concerning the results reported in the past \cite{Milanovic88,RF04,LCRP06}. We will refer to \textit{root-locus-like} terminology from now on throughout the paper, whenever we proceed to graph on the complex plane,  {the eigenvalues evolution for the non-linear problem as the band mixing parameter changes.} This complex-plane plot procedure, when a system's parameter varied, is amply known in the literature as the \emph{root locus} \cite{Evans48}. Taking advantage of the root-locus-like technique simulations \cite{AJGL11}, we were able to foretell features of the particle-scatterer interaction.

It is worthwhile to recall, there had also been several theoretical attempts to spread light to the effective potential dependency as $\kappa_{\textsc t}$ increases \cite{Ekbote99,Foreman07}. However, we do believe that ``monitoring" the quadratic eigenvalue problem \textit{via} the root-locus-like procedure \cite{AJGL11}, and plotting the metamorphosis of $V_{\mathrm{eff}}$, seem a promising way to get a better  {insight} into a complicated phenomenon  {referred to} as valence-band mixing. The last, is the target of the present theoretical study. In the next section, we present briefly the basic theoretical workbench to study $V_{\mathrm{eff}}$ in the valence band.  {In the following section, for highly specialized $III$-$V$ binary-compound semiconductor, the model is numerically tested for the evolution of $V_{\mathrm{eff}}$.} We devote the last section, to draft some conclusions.

\section{Basic theoretical formulation}
\label{sec:BasTeo}

An effective potential, is found useful to describe valence-band mixing in the EFA, \cite{Foreman07}. In the case envisioned here, to determine the operator {$\widehat{\bn{W}}_{\mathrm{eff}}$} for the effective \emph{band offset} potential, suffices to use the Kohn-L\"uttinger (KL) model Hamiltonian, \cite{KL55}, considering the transverse quasi-momentum $\vec{\kappa}_{\textsc t} = k_{x}\hat{e}_{x} + k_{y}\hat{e}_{y}$, because this is the direction of the Brillouin Zone where is described the present KL Hamiltonian. We assume understood any modification of the selected Brillouin Zone direction, as a change in the  {Hamiltonian model} to use. The system's quantal heterogeneity is considered along  {the} $z$ axis, taken perpendicular to the heterostructure interfaces. The operator  {$\widehat{\bn{W}}_{\mathrm{eff}}$}, is nothing but somewhat arbitrary convention, valid as long as one get holds of all potential-like energy terms from the original Hamiltonian operator, which are $z$-component momentum free \cite{Milanovic88,LCRP06}. Then

\begin{equation}
  \label{Weff}
   \widehat{\bn{W}}_{\mathrm{eff}} =
    \begin{bmatrix}
      W_{11} & W_{12} & 0 & 0 \\
      W_{12}^{*} &  W_{22} & 0 & 0 \\
      0 & 0 &  W_{22} & W_{12} \\
      0 & 0 & W_{12}^{*} & W_{11}
    \end{bmatrix},
  \end{equation}
\noindent is suitable for going through a standard calculation

\begin{equation}
  \label{Veff}
  \left[\widehat{\bn{W}}_{\mathrm{eff}}- V_{\mathrm{eff}}\bn{I}_{4}\right]\bn{\Psi}(z) = \bn{O}_{4},
\end{equation}
\noindent leading us to the effective potential \emph{band offset} $V_{\mathrm{eff}}$, ``\emph{felt}" in some sense, by holes during their passage trough the heterostructure, as $\kappa_{\textsc t}$ changes.

\begin{figure}
 \centering
 \includegraphics[width=3in,height=4in]{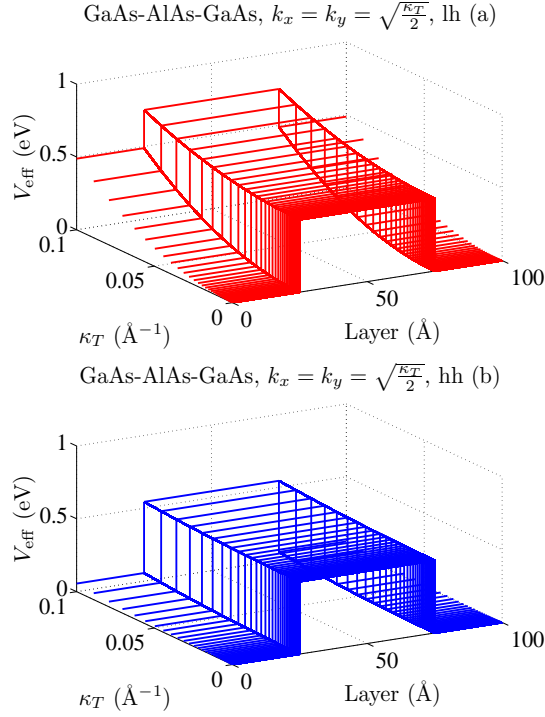}
   \caption{\label{F1-3D} Panel (a)/(b) displays the metamorphosis of the effective potential profile $V_{\mathrm{eff}}$ for \textit{lh}/\textit{hh} (red/blue lines), as a function of $\kappa_{\textsc t}$  and layer dimension for a \textsl{GaAs} {-}\textsl{AlAs}-\textsl{GaAs} heterostructure.}
\end{figure}

 {In the theoretical formulation, we use the periodic part of the Bloch functions whose components $\bn{u}_{0}(r) = (u_{1},u_{2},u_{3},u_{4})$}
\begin{eqnarray}
 \label{Basis}
 \left.
  \begin{array}{lcl}
    u_{1}  & = &  |\frac{3}{2}, \frac{3}{2}\rangle  =  \frac{1}{\sqrt{2}}
                                              |(x+iy)\uparrow\rangle \\
    u_{2} & = & |\frac{3}{2}, -\frac{1}{2}\rangle =  \frac{1}{\sqrt{6}}
      |(x-iy)\uparrow\rangle +\sqrt{\frac{2}{3}} |z\downarrow\rangle \\
    u_{3}  & = &  |\frac{3}{2}, \frac{1}{2}\rangle  =  \frac{1}{\sqrt{6}}
      |(x+iy)\downarrow\rangle - \sqrt{\frac{2}{3}} |z\uparrow\rangle \\
    u_{4}  & = &  |\frac{3}{2}, -\frac{3}{2}\rangle  =
     -\frac{i}{\sqrt{2}} |(x-iy)\downarrow\rangle
 \end{array}
\right\}\,,
\end{eqnarray}
\noindent  {were taken as basis functions to deal with the present KL model Hamiltonian} \cite{LCRP06}.  {The functions $|x\rangle, |y\rangle, |z\rangle\;$ have the lattice periodicity, and transform the atomic orbitals $p_{x}, p_{y}, p_{z}$, respectively. Here $|\uparrow\rangle, |\downarrow\rangle$, stand for the spin eigenfunctions. The order of (\ref{Basis}), follows that proposed in} Ref.\cite{Ikonic92b},  {which is:} $hh_{+\frac{3}{2}}, lh_{-\frac{1}{2}}, lh_{+\frac{1}{2}}, hh_{-\frac{3}{2}}$.

We have introduced
\begin{eqnarray*}
  W_{11} & = & A_{1}\kappa_{\textsc t}^{2} + V(z)\\
  W_{22} & = & A_{2}\kappa_{\textsc t}^{2} + V(z)\\
  W_{12} & = & \frac{\hbar^{2}\sqrt{3}}{2\,m_{0}}\left(\gamma_{2}(k_{y}^{2}-k_{x}^{2}) +
           2i\gamma_{3}k_{x}k_{y}\right)\\
  A_{1} & = & \frac{\hbar^{2}}{2m_{0}} (\gamma_{1}+\gamma_{2})\\
  A_{2} & = &  \frac{\hbar^{2}}{2m_{0}} (\gamma_{1}-\gamma_{2}),
\end{eqnarray*}
\noindent with $\gamma_{i}$ the L\"uttinger parameters, and $m_{0}$ the bare electron mass. The $\vec{\kappa}_{\textsc t}$ components $k_{x,y}$, are set in-plane respect to the heterostructure interfaces. In (\ref{Weff}), $\bn{I}_{4}/\bn{O}_4$ stands for the ($4 \times 4$) identity/null matrix, while in (\ref{Veff}), $\bn{\Psi}(z)$ is an envelope function. Though moderately rough, assertion (\ref{Veff}) represents a reliable-accuracy approximation to the $V_{\mathrm{eff}}$, we are interesting in. Let us consider a periodic three-layer [$A$-cladding left ($L$) layer /$B$ middle ($M$) layer/ $A$-cladding right ($R$) layer] heterostructure, in the absence of external fields or strains. In the bulk cladding layers, \emph{hh} and \emph{lh} modes mix due to the $\bn{k}\cdot\bn{p}$ interaction, while the middle slab represents a inhospitable medium for holes.  At zero valence-band mixing, one has

\begin{eqnarray}
  \label{Veff-0}
  \hspace{-6mm}\bn{V}(z) =
  \left\{
   \begin{array}{lcl}
     0 &;& \,\, z < z_{\textsc l}\\
     V_{\textsc b} - V_{\textsc a} = V_{\mathrm{eff}} &;& z_{\textsc l} < z <z_{\textsc r}\\
     0 &;& \,\, z > z_{\textsc r}
   \end{array}
   \right\}
   = \Theta V_{\mathrm{eff}},
\end{eqnarray}
\noindent being $\Theta$ a steplike function, and $V_{\textsc {a/b}}$ the potential of the cladding/middle layer.

\begin{figure}
 \centering
 \includegraphics[width=3in,height=4in]{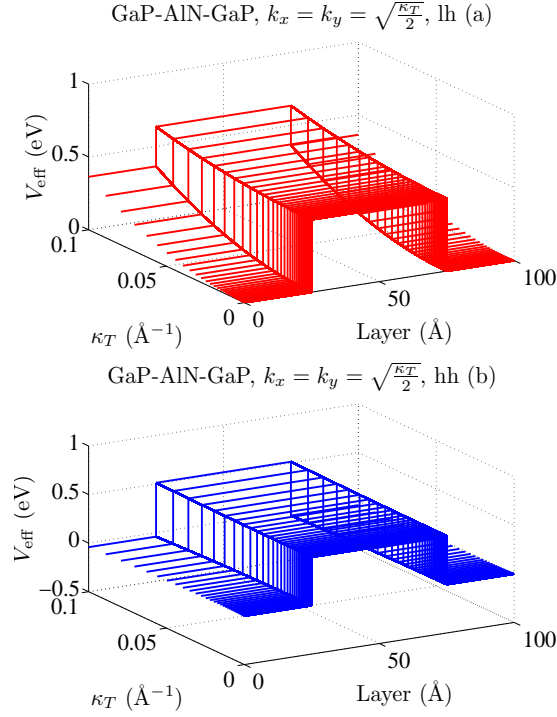}
   \caption{\label{F2-3D} Panel (a)/(b) presents the metamorphosis of the effective potential profile $V_{\mathrm{eff}}$ for \textit{lh}/\textit{hh} (red/blue lines), as a function of $\kappa_{\textsc t}$  and layer dimension for a \textsl{GaP}-\textsl{AlN}-\textsl{GaP} heterostructure.}
\end{figure}

Assuming a plane-wave-like dependence on $k_{z}$ for $\bn{\Psi}(z)$, and based on the quadratic eigenvalue problem (QEP) method,\cite{LCRP06} it can be cast

\begin{equation}
 \label{Polinom}
 \det\left[\bn{Q}(k_{z})\right] = q_{0} k_{z}^{8} + q_{1} k_{z}^{6} + q_{2} k_{z}^{4} + q_{3} k_{z}^{2} + q_{4},
\end{equation}

\noindent where $\bn{Q}(k_{z})$ is a second-degree matrix polynomial on the $z$-component wavevector $k_{z}$ and $q_{i}$ are function of L\"uttinger semi-empirical valance-band parameters and the components of $\kappa_{\textsc t}$. By dealing with (\ref{Polinom}), it is straightforward to follow whereas $k_{z}$ is oscillatory or not, and thereby the kind of $V_{\mathrm{eff}}$ the holes interplay with.

\begin{figure}
 \centering
 \includegraphics[width=3in,height=4in]{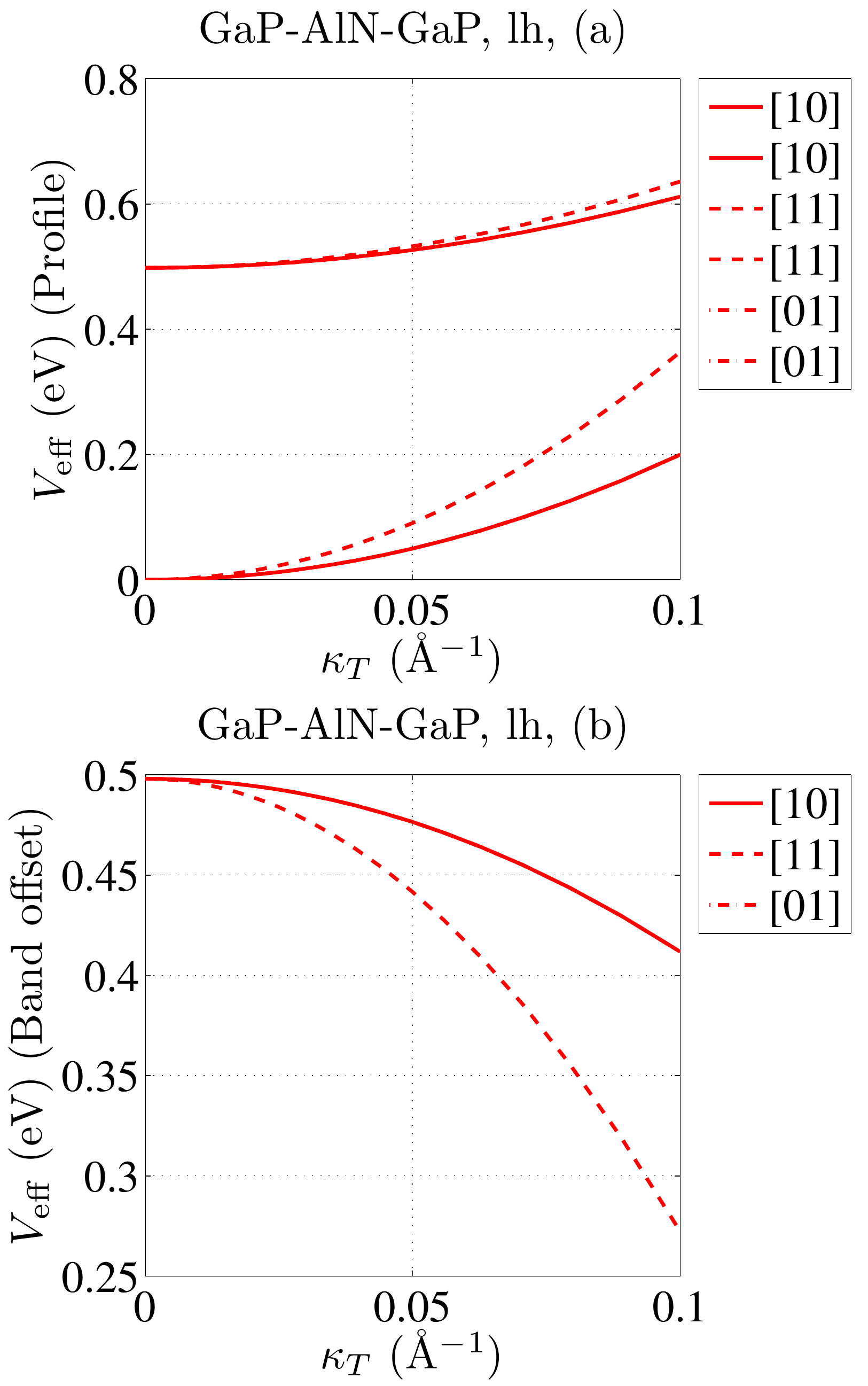}
 \includegraphics[width=3in,height=4in]{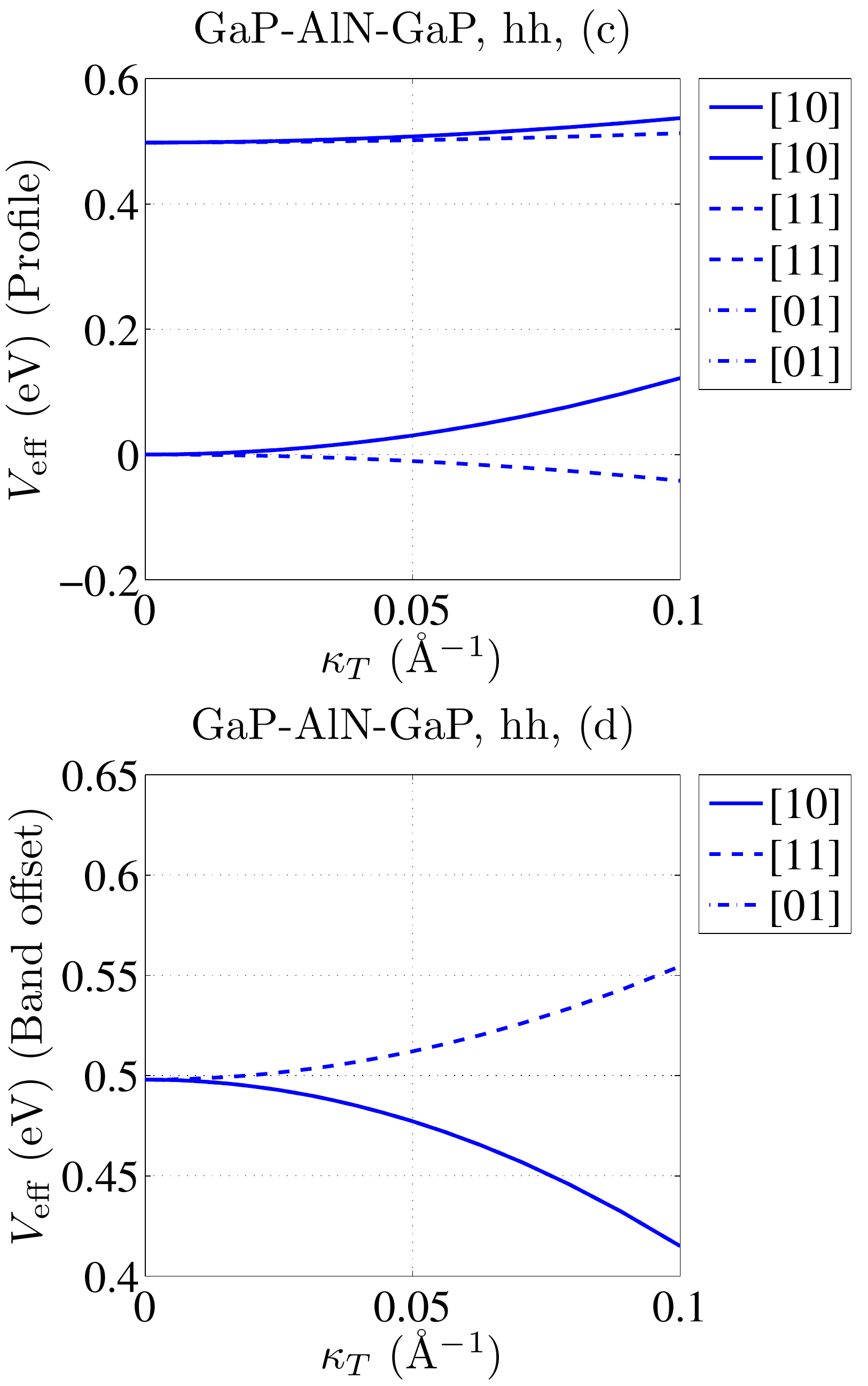}
   \caption{\label{F3-2D} Panel (a)/(c) displays a cut of the effective potential profile $V_{\mathrm{eff}}$ for \textit{lh}/\textit{hh} (red/blue lines), at the interface plane between left and middle layers, as a function of $\kappa_{\textsc t}$.  Panel (b)/(d) shows the progression of the \emph{band offset} at the same interface for \textit{lh}/\textit{hh} (red/blue lines), \textit{i.e.} the difference between the upper-edge and lower-edge of the $V_{\mathrm{eff}}$ profile. We have considered a \textsl{GaP}-\textsl{AlN}-\textsl{GaP} layered heterostructure.  {Notice the coincidence of the curves along [10] (solid line) and [01] (dashed-dotted line), in-plane directions.}}
\end{figure}

\section{Numerical results and discussion}
\label{sec:SimDis}

On general grounds, for $\kappa_{\textsc t} \approx 0$ the effective \emph{band offset} is given by the difference for $3D$ band edge levels, \cite{LCRP06,RF04} By letting grow $\kappa_{\textsc t}$, the band mixing effects arise and the effective \emph{band offset} changes. A comprehensive analysis of this subject, describes larger reduction for the piecewise constant effective barrier height $V_{\mathrm{eff}}$ as a function of $\kappa_{\textsc t}$, for \emph{lh} respect to that for \emph{hh} \cite{LCRP06}. The mechanism responsible for this behavior is the increment of the term $\kappa_{\textsc t}^{2}/m^{*}_{hh,lh}(z)$, yielding even to invert the roles of wells and barriers \cite{RF04,Milanovic88} as can  {be} derived from equation (\ref{Polinom}). Some authors had declared a shift upward in energy, of the bound states in the effective potential well as the transverse wave vector increases,\cite{Ekbote99}. Recently a valence-band mixing first-principle theory within the EFA was proposed, which approximated the superlattice potential energy by considering only the linear and quadratic responses to the heterostructure perturbation\cite{Foreman07}.

Today the $III$-$V$ binary- and ternary-compound semiconductor alloys, continue attracting interest, for real-life applications and facilities. Nano-dimensional systems based on $AlAs$, $AlSb$, and $AlGaN$, together with quantum wells composed on $GaAs$, $InAs$, $GaP$, $GaSb$ and $GaN$ components, are a robust platform for material science, condensed-matter physics and for a rapidly raising field of Spintronics \cite{Loss09}. In this concern, we briefly describe here some phenomenological properties of several representative $III$-$V$ binary-compound semiconductor alloys. To gain some insight into the rather complicated influence of the band mixing parameter $\kappa_{\textsc t}$, on the effective \emph{band offset}, we display several graphics in the present section. The central point here, is a reliable numerical simulation for the spatial distribution of $V_{\mathrm{eff}}$ while the valence-band mixing increases from $\kappa_{\textsc t} \approx 0$ (uncoupled holes) to $\kappa_{\textsc t} = 0.1$\AA$^{-1}$ (strong hole band mixing). This purpose requires a solution of (\ref{Veff}) looking for a systematic start-point theoretical treatment of highly specialized $III$-$V$ semiconductor binary-compound cases of interest. We have set a width of $25\,$\AA$\;$ for the external cladding-layer $L$ and $R$, while for the middle one  we have taken a thickness of $50\,$\AA.

\begin{figure}
 \centering
 \includegraphics[width=3in,height=4in]{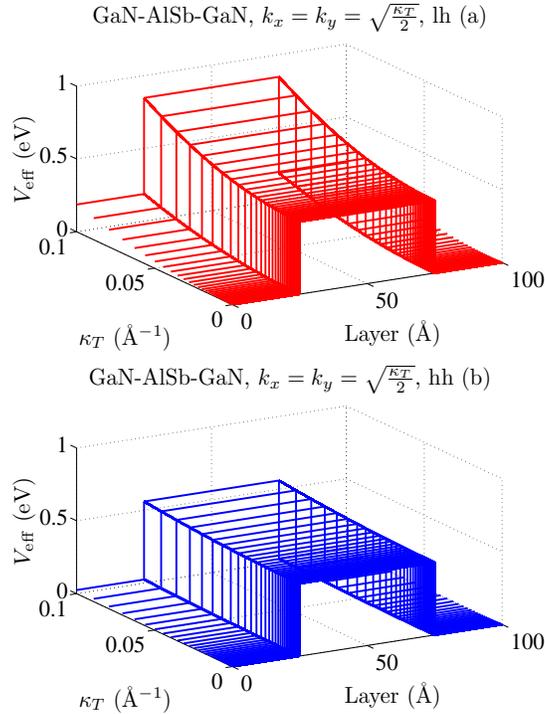}
   \caption{\label{F4-3D} Panel (a)/(b) displays the metamorphosis of the effective potential profile $V_{\mathrm{eff}}$ for \textit{lh}/\textit{hh} (red/blue lines), as a function of $\kappa_{\textsc t}$  and layer dimension for a \textsl{GaN}-\textsl{AlSb}-\textsl{GaN} heterostructure.}
\end{figure}

Figure \ref{F1-3D} demonstrates that the standard rectangular distribution for $V_{z}$ (\ref{Veff-0}), is a consistent potential-energy trial for \textit{hh} (blue lines), applicable in the wide range of $\kappa_{\textsc t}$ [see panel (b)]. On the contrary, panel (a) remarks that the rectangular shape for $V_{z}$ is no longer valid for \textit{lh} (red lines), as $\kappa_{\textsc t}$ increases.  {It is} in this very sense, when the valence-band mixing effects get rise, that become unavoidable to refer an effective \emph{band offset} for a realistic description of the interplay of the envisioned physical structure with holes. We display in panel (a), the metamorphosis of $V_{\mathrm{eff}}$ for \textit{lh}, as a function of $\kappa_{\textsc t}$  and layer dimension for a \textsl{GaAs}-\textsl{AlAs}-\textsl{GaAs} heterostructure. Two changes are neatly observable, namely: the energy edge of both left and right cladding-layers steps up in almost $0.5$ eV, while for the middle one it remains almost constant. The last departs from the $V_{\mathrm{eff}}$ evolution for \textit{hh}, where all borders move up almost rigidly [see panel (b)]. {Let us, briefly comment these results in an attempt of a qualitative comparison with ones given in} \cite{Ikonic92b},  {when a finding of hole states for a quantum well  embedded in asymptotically flat potential was addressed}.  {We conjecture the obtained by them dispersion of lower-\textit{hh} bound levels} \cite{Ikonic92b},  {to be considered reasonably correct even for larger $\kappa_{\textsc t}$, due to the quasi-flat profile of $V_{\mathrm{eff}}$, observed for \textit{hh} in our Fig.}\ref{F1-3D}(b),  {at any value of band mixing}.  {While for \textit{lh} states, some changes could be expected in the lower bound states} \cite{Ikonic92b},  {as the effective barrier height ---for \textit{lh}---, is no longer fixed, but rather becomes a dynamically modifiable quantity with $\kappa_{\textsc t}$, as can be seen from our Fig.}\ref{F1-3D}(a). Although not shown here owing to brevity, a similar behavior was found for other middle-layer alloys (\textsl{AlSb}, \textsl{AlP}, \textsl{AlN}). Described above features, remain under modification of the in-plane direction.

\begin{figure*}
 \centering
 \includegraphics[width=5.5in,height=4in]{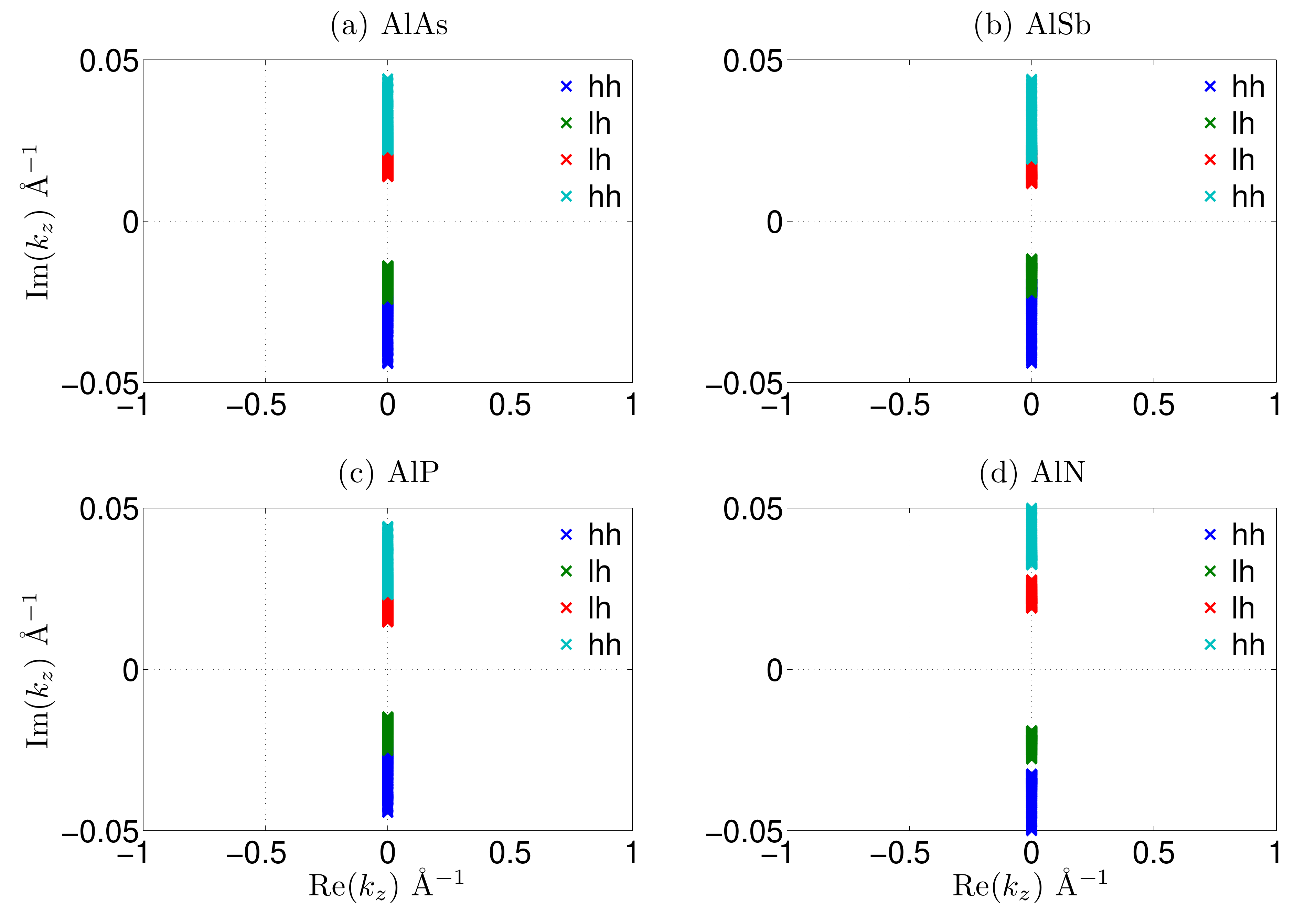}
  \caption{\label{F5-spectra} Root-locus-like map for the eigenvalues $k_{z}$ from QEP's solutions (\ref{Polinom}), as a function of $k_{x}=k_{y}$ with $\kappa_{\textsc t} \in \left[10^{-6},10^{-1} \right]$ \AA$^{-1}$ for: (a) \textsl{AlAs}, (b) \textsl{AlSb}, (c) \textsl{AlP} and (d) \textsl{AlN} binary compounds. We had assumed $V(z)=0.498$ eV, and the longitudinal energy $E = 0.475$ eV. The map evolves outward coordinate center of imaginary axis, with different values of $k_{z}$ for both \emph{lh} and \emph{hh}.}
\end{figure*}

Figures \ref{F2-3D} and \ref{F3-2D} illustrate the phenomenology in another group of $III$-$V$ semiconductor binary-compounds. They show the $V_{\mathrm{eff}}$ metamorphosis for a \textsl{GaP}-\textsl{AlN}-\textsl{GaP} layered heterostructure. In panel (a) of Fig.\ref{F2-3D}, we observed a reduction of the effective \emph{band offset} for \textit{lh} (red lines), similar to that of a \textsl{GaAs}-\textsl{AlAs}-\textsl{GaAs} heterostructure [see Fig.\ref{F1-3D}(a)]. However, the panel (b) of Fig.\ref{F2-3D} illustrates a particularly different behavior for \textit{hh} (blue lines), due to a decrease of the energy edge for both left and right cladding-layers. A more explicit representation of that, is shown in Fig.\ref{F3-2D}(c). {Concerning  results of all panels in Fig.}\ref{F3-2D},  {we found useful to underline the coincidence for $V_{\mathrm{eff}}$ profile as well as for \emph{band offset} along [10] (solid line) and [01] (dashed-dotted line), in-plane directions. The last is a clear evidence of a partial isotropic behavior for the overlapped lines}.  {From Fig.}\ref{F3-2D}(c), it can be noticed, how $V_{\mathrm{eff}}$ diminishes its lower-edge profile down to $-0.05$ eV roughly, along the [11] in-plane direction (blue dashed line). Meanwhile, the upper-edge profile evolution plotted in Fig.\ref{F3-2D}(c), goes on nearly constant along the same direction. On the other hand, in this kind of heterostructure, the in-plane full-isotropic performance of $V_{\mathrm{eff}}$ evolution, is absence.  This fact is observed from panels (c)-(d) of Fig.\ref{F3-2D}.  {However}, it can be assumed again a rectangular-like distribution of $V_{z}$ (\ref{Veff-0}) for \textit{hh} (blue lines), as an  {acceptable} reference frame along the [10] (solid line) and [01] (dashed-dotted line), in-plane directions. Moreover, notice that the difference between profile edges [see Fig.\ref{F3-2D}(d)] is less than $0.1$ eV at the higher $\kappa_{\textsc t}$ value, respect to the zero hole-mixing point. Concerning the same in-plane directions, worthwhile to underline that for \textit{lh}, a slightly reduced $V_{\mathrm{eff}}$ profile progression, stays as trend [see Fig.\ref{F3-2D}(a), red solid line]. Notice at panel Fig.\ref{F3-2D}(b), that the $V_{\mathrm{eff}}$ \emph{band offset} evolves up to $0.4$ eV along [10] and [01] in-plane directions at maximum $\kappa_{\textsc t}$, while it reduces almost to $0.2$ eV for the [11] interface orientation.

Figure \ref{F4-3D} plots the evolution of $V_{\mathrm{eff}}$ for a \textsl{GaN}-\textsl{AlSb}-\textsl{GaN} layered heterostructure. In the presence of nitrides in the external layers, the noticeable increment of the effective \emph{band offset} for \textit{lh} (red lines) observed in panel (a) of Fig.\ref{F4-3D}, contrasts with the reduction we had shown in panel (a) of Fig.\ref{F3-2D} for non-nitride alloys. Meanwhile in this case, the \textit{hh} (blue lines) feel a similar distribution as in \textsl{GaAs}-\textsl{AlAs}-\textsl{GaAs} heterostructure (see Figure \ref{F1-3D}).

For the sake of illustration, we retrieve a procedure that relies on a robust classical control theory technique, a so-called root locus \cite{Evans48}. Once we have quoted the eigenvalues $k_z$ of (\ref{Polinom}), it is then straightforward to generate a plot in the complex plane, symbolizing the  {locations} of $k_z$ values that rise as a band mixing parameter $\kappa_{\textsc t}$ changes. We take advantage of the root-locus-like know-how, to promptly identify evanescent modes, keeping in mind that complex (or pure imaginary) solutions are displayed in the up(down)-half plane (see Figure \ref{F5-spectra}). These values are forbidden for some layers and represent unstable solutions, underlining the lack of hospitality of these slabs for oscillating modes. Although not shown here, the opposite examination is also suitable for propagating modes, which become patterned in left(right)-half plane, and are equated with stable solutions for given  layers. Thus we are able ``to stamp'' on a $2D$-map language, the stability domain analysis of the envisioned heterostructure under a quantum-transport problem. This way, we are presenting an unfamiliar methodology to deal with solid-state low-dimensional physical phenomenology.

\section{Conclusions}
\label{sec:Conclu}
We present an alternative procedure to simulate graphically, the phenomenon of the transverse degree of freedom dependence of the effective scattering potential. For low-intensity valence-band mixing regime, a fixed-height rectangular distribution of the potential-energy, is a good trial as a standard reference frame, for a theoretical treatment involving both flavors of holes in the materials under study. However, this assertion is no longer valid for light holes, whenever the mixing effects reveal, and a mutable effective \emph{band offset} profile has to be taken into account, whenever the in-plane kinetic energy changes. Meanwhile, for heavy holes the effective \emph{band offset} profile assumed at $\kappa_{\textsc t} \approx 0$ remains robust for finite values of valence-band mixing. Evidences of this sort, are relevant in experimental applications, and in theoretical analysis of coupled-hole tunneling.

\section*{Acknowledgments}
This work was developed under support of FICSAC, UIA, M\'{e}xico. One of the authors (L.D-C) is grateful to the Visiting Academic Program of the UIA, M\'{e}xico.

\end{document}